\documentclass[12pt]{article}

\usepackage{epsfig}

{\end{list}}
\newcounter{enumct}

\begin{document}

\sloppy

\pagestyle{empty}

\newcommand{\G}{\gamma}
\newcommand{\GG}{{\gamma\gamma}}
\newcommand{\EPEM}{e^+e^-}
\newcommand{\MUPM}{\mu^+\mu^-}
\newcommand{\A}{\alpha}
\newcommand{\ZA}{Z\alpha}
\newcommand{\BE}{\begin{equation}}
\newcommand{\EE}{\end{equation}}

\makeatletter
\def\lesssim{\mathrel{\mathpalette\vereq<}}
\def\vereq#1#2{\lower3pt\vbox{\baselineskip1.5pt \lineskip1.5pt
\ialign{$\m@th#1\hfill##\hfil$\crcr#2\crcr\sim\crcr}}}
\def\gtrsim{\mathrel{\mathpalette\vereq>}}
\makeatother

\input{boxedeps.tex}
\SetRokickiEPSFSpecial
\HideDisplacementBoxes


\begin{center}
{\LARGE\bf Two Photon Physics in pp and AA Collisions}\\[10mm]
{\Large Gerhard Baur} \\[3mm]
{\it Institut f\"ur Kernphysik,}\\[1mm]
{\it Forschungszentrum J\"ulich, J\"ulich, Germany}\\[1mm]
{\it E-mail: G.Baur@fz-juelich.de}\\[5mm]
{\Large Kai Hencken} \\[3mm]
{\it Institut f\"ur Physik,}\\[1mm]
{\it Universit\"at Basel, Basel, Switzerland}\\[1mm]
{\it E-mail: hencken@quasar.physik.unibas.ch}\\[5mm]
{\Large Dirk Trautmann} \\[3mm]
{\it Institut f\"ur Physik,}\\[1mm]
{\it Universit\"at Basel, Basel, Switzerland}\\[1mm]
{\it E-mail: trautmann@ubaclu.unibas.ch}\\[20mm]

{\bf Abstract}\\[1mm]
\begin{minipage}[t]{140mm}

In central collisions at relativistic heavy ion colliders like the
Relativistic Heavy Ion Collider RHIC/Brookhaven and the Large Hadron
Collider LHC (in its heavy ion mode) at CERN/Geneva, one aims at
detecting a new form of hadronic matter --- the Quark Gluon Plasma.
We discuss here a complementary aspect of these collisions, the very
peripheral ones. Due to coherence, there are strong electromagnetic
fields of short duration in such collisions. They give rise to
photon-photon and photon-nucleus collisions with high flux up to an
invariant mass region hitherto unexplored experimentally. After a
general survey photon-photon luminosities in relativistic heavy ion
collisions are discussed. Special care is taken to include the effects
of strong interactions and nuclear size. Then photon-photon physics at
various $\GG$-invariant mass scales is discussed.  The region of
several GeV, relevant for RHIC is dominated by QCD phenomena (meson
and vector meson pair production). Invariant masses of up to about 100
GeV can be reached at LHC, and the potential for new physics is
discussed. Photonuclear reactions and other important background
effects, especially diffractive processes are also discussed.
Lepton-pair production, especially electron-positron pair production
is copious. Due to the strong fields there will be new phenomena,
especially multiple $\EPEM$ pair production.

\end{minipage}\\[5mm]

\rule{160mm}{0.4mm}

\end{center}

\section{Introduction}

The parton model is very useful to study scattering processes at very
high energies. The scattering is described as an incoherent
superposition of the scattering of the various constituents. For
example, nuclei consist of nucleons which in turn consist of quarks
and gluons, photons consist of lepton pairs, electrons consist of
photons, etc.. We note that relativistic nuclei have photons as an
important constituent, especially for low enough virtuality
$Q^2=-q^2>0$ of the photon. This is due to the coherent action of all
the charges in the nucleus.  The virtuality of the photon is related
to the size $R$ of the nucleus by
\BE
Q^2 \lesssim 1/R^2, 
\EE
the condition for coherence. The radius of a nucleus is given
approximately by $R=1.2$~fm~$A^{1/3}$, where $A$ is the nucleon
number. From the kinematics of the process one has
\BE
Q^2=\frac{\omega^2}{\G^2}+q_\perp^2.
\EE
Due to the coherence condition the maximum energy of the
quasireal photon is therefore given by
\BE
\omega_{max} \approx \frac{\G}{R},
\label{eq_wmax}
\EE
and the maximum value of the perpendicular component is given by
\BE
q_\perp \lesssim \frac{1}{R}.
\EE
We define the
ratio $x=\omega/E$, where $E$ denotes
the energy of the nucleus $E= M_N \G A$ and $M_N$ is the nucleon mass.
It is therefore smaller than
\BE
 x_{max}=\frac{1}{R M_N A} = \frac{\lambda_C(A)}{R},
\EE
where $\lambda_C(A)$ is the Compton wave length of the ion. Here and
also throughout the rest of the paper we use natural units, setting
$\hbar=c=1$.

The collisions of $e^+$ and $e^-$ has been the traditional way to
study $\GG$-collisions. Similarly photon-photon collisions can also be
observed in hadron-hadron collisions. Since the photon number scales
with $Z^2$ ($Z$ being the charge number of the nucleus) such effects
can be particularly large. Of course, the strong interaction of the
two nuclei has to be taken into consideration.
%
%
\begin{figure}[tbh]
\begin{center}
\ForceHeight{4cm}
\BoxedEPSF{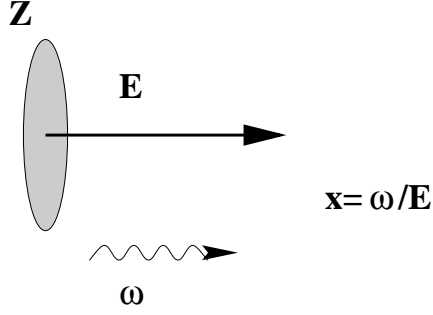}
\end{center}
\caption{\it
A fast moving nucleus with charge $Ze$ is surrounded by a strong
electromagnetic field. This can be viewed as a cloud of virtual
photons.  These photons can often be considered as real. They are
called equivalent or quasireal photons. The ratio of the photon energy
$\omega$ and the incident beam energy $E$ is denoted by $x=\omega/E$.
Its maximal value is restricted by the coherence condition to
$x<\lambda_C(A)/R\approx 0.175/A^{4/3}$, that is, $x\protect\lesssim
10^{-3}$ for Ca ions and $x\protect\lesssim 10^{-4}$ for Pb ions.  }

\label{fig_xvar}
\end{figure}
%
%
\begin{figure}[tbh]
\begin{center}
\ForceHeight{5cm}
\BoxedEPSF{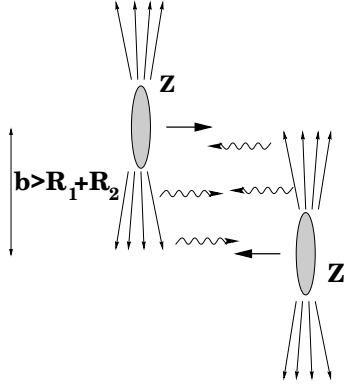}
\end{center}
\label{fig_collision}
\caption{\it
Two fast moving electrically charged objects are an abundant source of
(quasireal) photons. They can collide with each other and with the
other nucleus. For peripheral collisions with impact parameters $b>2R$,
this is useful for photon-photon as well as photon-nucleus
collisions.}
\end{figure}

The equivalent photon flux present in medium and high energy nuclear
collisions is very high, and has found many useful applications in
nuclear physics \cite{BertulaniB88}, nuclear astrophysics
\cite{BaurR94,BaurR96}, particle physics \cite{Primakoff51} (sometimes
called the ``Primakoff effect''), 
as well as, atomic physics \cite{Moshammer97}.
It is the main purpose of this review to discuss the physics of
photon-photon and photon-hadron (nucleus) collisions in high energy
heavy ion collisions. With the 
construction of the ``Relativistic Heavy Ion Collider'' (RHIC) and the
``Large Hadron Collider'' (LHC) scheduled for 1999 and for 2004/2008,
respectively, one will be able to investigate such collisions
experimentally. The main purpose of these heavy ion colliders 
is the formation and detection of the
quark-gluon-plasma, a new form of highly excited dense hadronic
matter. Such a state of matter will be created in central
collisions. The present interest is in the ``very peripheral (distant)
collisions'', where the nuclei do not interact strongly with each
other. From this point of view, grazing collisions and central
collisions are considered as a background. It is needless to say that
this ``background'' can also be interesting physics of its own.

The equivalent photon spectrum extends up to several GeV at RHIC
energies ($\G\approx 100$) and up to about 100 GeV at LHC energies
($\G\approx 3000$), see Eq.~(\ref{eq_wmax}).  Therefore the range of
invariant masses $M_{\GG}$ at RHIC will be up to about the mass of the
$\eta_c$, at LHC it will extend into an invariant mass range hitherto
unexplored.

Relativistic heavy ion collisions have been suggested as a general
tool for two photon physics about a decade ago. Yet the study of a
special case, the production of $\EPEM$ pairs in nucleus-nucleus
collisions, goes back to the work of Landau and Lifschitz in 1934
\cite{LandauL34} (In those days, of course, one thought more about
high energy cosmic ray nuclei than relativistic heavy ion
colliders).
The general possibilities and characteristic features of two-photon
physics in relativistic heavy ion collisions have been discussed in
\cite{BaurB88}. The possibility to produce a Higgs boson via
$\GG$-fusion was suggested in \cite{GrabiakMG89,Papageorgiu89}. In
these papers the effect of strong absorption in heavy ion collisions
was not taken into account. This absorption is a feature, which is
quite different from the two-photon physics at $\EPEM$ colliders. The
problem of taking strong interactions into account was solved by using
impact parameter space methods in \cite{Baur90d,BaurF90,CahnJ90}. Thus
the calculation of $\GG$-luminosities in heavy ion collisions is put
on a firm basis and rather definite conclusions were reached by many
groups working in the field \cite{VidovicGB93}; for recent reviews 
see \cite{KraussGS97} and \cite{BaurHT98}. This opens the way
for many interesting applications.
Up to now hadron-hadron collisions have not been used for
two-photon physics. An exception can be found in
\cite{Vannucci80}. There the production of $\mu^+\mu^-$ pairs at the
ISR was observed.  The special class of events was selected, where no
hadrons are seen associated with the muon pair in a large solid angle
vertex detector. In this way one makes sure that the hadrons do not
interact strongly with each other, i.e., one is dealing with
peripheral collisions (with impact parameters $b>2R$); the
photon-photon collisions manifest themselves as ``silent events''.
Dimuons with a very low sum of transverse momenta are also considered
as a luminosity monitor for the ATLAS detector at LHC \cite{ags98}.

Experiments are planned at RHIC
\cite{KleinS97a,KleinS97b,KleinS95a,KleinS95b,jny98} and discussed at
LHC \cite{HenckenKKS96,Felix97,BaurHTS98}.
We quote J. D. Bjorken \cite{Bjorken99}:
{\it It is an important portion (of the FELIX program at LHC)to tag on
Weizsaecker Williams photons(via the nonobservation of completely
undissociated forward ions)in ion-ion running, creating a high
luminosity $\gamma-\gamma$collider.}

\section{From impact-parameter dependent equivalent photon spectra to
{$\GG$-luminosities}}
\label{sec_lum}

\begin{figure}[tbhp]
\begin{center}
\ForceHeight{5cm}
\BoxedEPSF{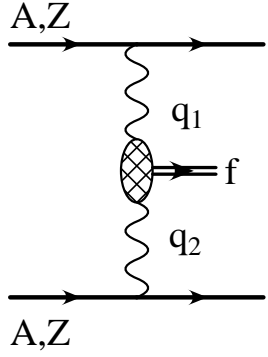}
\end{center}
\caption{\it
The general Feynman diagram of photon-photon processes in heavy ion
collisions: Two (virtual) photons fuse in a charged particle collision
into a final system $f$.
} 
\label{fig_ggcollision}
\end{figure}

Photon-photon collisions have been studied extensively at $\EPEM$
colliders. The theoretical framework is reviewed, e.g., in
\cite{BudnevGM75}.  The basic graph for the two-photon process in
ion-ion collisions is shown in Fig.~\ref{fig_ggcollision}. Two virtual
(space-like) photons collide to form a final state $f$. In the
equivalent photon approximation it is assumed that the square of the
4-momentum of the virtual photons is small, i.e., $q_1^2\approx
q_2^2\approx 0$ and the photons can be treated as quasireal. In this
case the $\GG$-production is factorized into an elementary cross
section for the process $\G+\G\rightarrow f$ (with real photons, i.e.,
$q^2=0$) and a $\GG$-luminosity function. In contrast to the pointlike
elementary electrons (positrons), nuclei are extended, strongly
interacting objects with internal structure. This gives rise to
modifications in the theoretical treatment of two photon processes.
The emission of a photon depends on the (elastic) form factor. Often a
Gaussian form factor or one of a homogeneous charged sphere is used.
The typical behavior of a form factor is
\BE
f(q^2) \approx 
\left\{
 \begin{array}{lcl}
 Z &\qquad& \mbox{for $|q^2| < \frac{1}{R^2}$}\\
 0 &\qquad& \mbox{for $|q^2| \gg \frac{1}{R^2}$}
 \end{array}
\right. .
\EE
For low $|q^2|$ all the protons inside the nucleus act coherently,
whereas for $|q^2| \gg 1/R^2$ the form factor is very small, close to
0. For a medium size nucleus with, say, $R=5$ fm, the limiting
$Q^2=-q^2=1/R^2$ is given by $Q^2=(40$MeV$)^2=1.6\times
10^{-3}$~GeV${}^2$. Apart from $\EPEM$ (and to a certain extent also
$\mu^+\mu^-$) pair production, this scale is much smaller than typical
scales in the two-photon processes. Therefore the virtual photons in
relativistic heavy ion collisions can be treated as quasireal. This is
a limitation as compared to $\EPEM$ collisions, where the two-photon
processes can also be studied as a function of the corresponding
masses $q_1^2$ and $q_2^2$ of the exchanged photon (``tagged mode'').
%
%

As was discussed already in the previous section, relativistic
heavy ions interact strongly when the impact parameter is smaller than
the sum of the radii of the two nuclei. In such cases $\GG$-processes
are still present and are a background that has to be considered in
central collisions. In order to study ``clean'' photon-photon events
however, they have to be eliminated in the calculation of
photon-photon luminosities as the particle production due to the
strong interaction dominates. In the usual treatment of photon-photon
processes in $\EPEM$ collisions plane waves are used and there is
no direct information on the impact parameter. For heavy ion
collisions on the other hand it is very appropriate to introduce
impact parameter dependent equivalent photon numbers. They have been
widely discussed in the literature, see, e.g.,
\cite{BertulaniB88,JacksonED,WintherA79}.

The equivalent photon spectrum corresponding to a point charge $Z e$,
moving with a velocity $v$ at impact parameter $b$ is given by
\BE
N(\omega,b) = \frac{Z^2\A}{\pi^2} \frac{1}{b^2}
\left(\frac{c}{v}\right)^2 x^2 \left[ K_1^2(x) + \frac{1}{\G^2}
K_0^2(x)\right],
\label{eq_nomegab}
\EE
where $K_n(x)$ are the modified Bessel Functions (MacDonald
Functions) and $x=\frac{\omega b}{\G v}$. 
Then one obtains the probability for a certain electromagnetic process
to occur in terms of the same process generated by an equivalent pulse
of light as
\BE
P(b) = \int \frac{d\omega}{\omega} N(\omega,b) \sigma_\G(\omega).
\EE
Possible modifications of $N(\omega,b)$ due to an extended spherically
symmetric charge distribution are given in \cite{BaurF91}. It should
be noted that Eq.~(\ref{eq_nomegab}) also describes the equivalent
photon spectrum of an extended charge distribution, such as a nucleus,
as long as $b$ is larger than the extension of the object. This is due
to the fact that the electric field of a spherically symmetric system
depends only on the total charge, which is inside it. As one often
wants to avoid also final state interaction between the produced
system and the nuclei, one has to restrict oneself to $b_i>R_i$ and
therefore the form factor is not very important.

As the term $x^2 \left[ K_1^2(x) + 1/\G^2 K_0^2(x)\right]$ in
Eq.~(\ref{eq_nomegab}) can be roughly approximated as 1 for $x<1$ and
0 for $x>1$, so that the equivalent photon number
$N(\omega,b)$ is almost a constant up to a maximum
$\omega_{max}=\G/b$ ($x=1$). By integrating
the photon spectrum (Eq.~(\ref{eq_nomegab}))
over $b$ from a minimum value of $R_{min}$ up to infinity (where
essentially only impact parameter up to  $b_{max}\approx \G/\omega$ 
contribute, compare with Eq.~(\ref{eq_wmax})),
one can define an equivalent photon number
$n(\omega)$. This integral can be carried out analytically and is
given by \cite{BertulaniB88,JacksonED}
\BE
n(\omega) = \int d^2b N(\omega,b) = \frac{2}{\pi} Z_1^2 \alpha
\left(\frac{c}{v}\right)^2 \left[ \xi K_0 K_1 - \frac{v^2\xi^2}{2 c^2}
\left(K_1^2 - K_0^2\right)\right] ,
\label{eq_nomegaex}
\EE
where the argument of the modified Bessel functions is 
$\xi=\frac{\omega R_{min}}{\G v}$.
The cross section for a certain electromagnetic process is then
\BE
\sigma = \int \frac{d\omega}{\omega} n(\omega) \sigma_{\G}(\omega).
\label{eq_sigmac}
\EE
Using the approximation above for the MacDonald functions, we get
an approximated form, which is quite reasonable and is useful for
estimates:
\BE
n(\omega) \approx  \frac{2 Z^2 \A}{\pi} \ln
\frac{\G}{\omega R_{min}}.
\label{eq_nomegaapprox}
\EE

The photon-photon production cross-section is obtained in a similar
factorized form,
by folding the corresponding equivalent photon spectra of the two
colliding heavy ions \cite{BaurF90,CahnJ90} (for
polarization effects see \cite{BaurF90}, they are neglected here)
\BE
\sigma_c = \int \frac{d\omega_1}{\omega_1} \int
\frac{d\omega_2}{\omega_2}
F(\omega_1,\omega_2) \sigma_{\GG}(W_{\GG}=\sqrt{4 \omega_1\omega_2}) ,
\label{eq_sigmaAA}
\EE
with
\begin{eqnarray}
F(\omega_1,\omega_2)&=& 2\pi \int_{R_1}^{\infty} b_1 db_1 
\int_{R_2}^{\infty} b_2 db_2 \int_0^{2\pi} d\phi \nonumber\\
&&\times N(\omega_1,b_1) N(\omega_2,b_2)
\Theta\left(b_1^2+b_2^2-2 b_1 b_2
\cos\phi-R_{cutoff}^2\right) ,
\label{eq_fw1w2}
\end{eqnarray}
($R_{cutoff} = R_1 + R_2$).  This can also be rewritten in terms of
the invariant mass $W_{\GG}=\sqrt{4\omega_1\omega_2}$ and the rapidity
$Y=1/2 \ln[(P_0+P_z)/(P_0-P_z)]=1/2 \ln(\omega_1/\omega_2)$ as:
\BE
\sigma_c = \int dW_{\GG} dY \frac{d^2L}{dW_{\GG} dY}
\sigma_{\GG}(W_{\GG}) ,
\label{eq_sigmaAAMY}
\EE
with 
\BE
\frac{d^2L_{\GG}}{dW_{\GG} dY} = \frac{2}{W_{\GG}}
F\left(\frac{W_{\GG}}{2} e^Y,\frac{W_{\GG}}{2} e^{-Y}\right) .
\label{eq_dldwdy}
\EE
Here energy and momentum in the beam direction are
denoted by $P_0$ and $P_z$. The transverse momentum is of the order of
$P_\perp \le 1/R$ and is neglected here. The transverse momentum
distribution is calculated in \cite{BaurB93}.

In \cite{BaurB93} and \cite{Baur92}
this intuitively plausible formula is derived ab
initio, starting from the assumption that the two ions move on a
straight line with impact parameter $b$. Eqs.~(\ref{eq_sigmaAA})
and~(\ref{eq_sigmaAAMY}) are the basic formulae for $\GG$-physics in
relativistic heavy-ion collisions. The advantage of heavy nuclei is
seen in the coherence factor $Z_1^2 Z_2^2$ contained in
Eqs.~(\ref{eq_sigmaAA})--(\ref{eq_dldwdy}).

As a function of $Y$, the luminosity $d^2L/dW_{\GG}{dY}$ has a
Gaussian shape with the maximum at $Y=0$. The width is approximately
given by $\Delta Y = 2 \ln \left[(2\G)/(R W_{\GG})\right]$. Depending
on the experimental situation additional cuts in the allowed $Y$ range 
are needed.

Additional effects due to the nuclear structure have been also studied.
For inelastic vertices a photon number $N(\omega,b)$ can also be defined, 
see, e.g., \cite{BaurHT98}. Its effect was found to be small. The
dominant correction comes from the electromagnetic excitation of one
of the ion in addition to the photon emission. 
We refer to \cite{BaurHT98} for further details.

\begin{figure}[tbh]
\begin{center}
\ForceHeight{7.5cm}
\BoxedEPSF{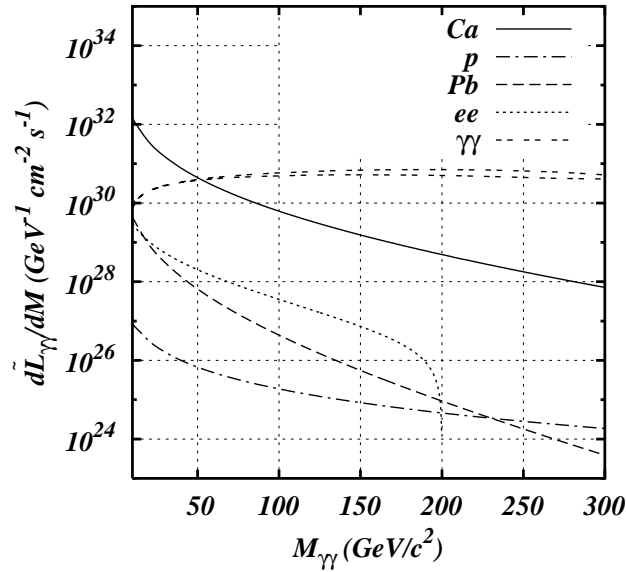}
\end{center}
\caption{\it Comparison of the effective $\GG$-Luminosities ($L_{AA}
\times L_{\GG}$) for different ion species. For comparison the same
quantity is shown for LEP200 and a future NLC/PLC (next linear
collider/photon linear collider,photons are obtained by LASER backscattering, two different polarizations are
shown).}
\label{fig_lum}
\end{figure}

In Fig.~\ref{fig_lum}
we give a comparison of effective $\GG$ luminosites for various 
collider scenarios.
We use the following collider parameters: LEP200: $E_{el}=100$GeV,
$L=10^{32} cm^{-2} s^{-1}$, NLC/PLC: $E_{el}=500$GeV, 
$L=2 \times 10^{33} cm^{-2} s^{-1}$, Pb-Pb heavy-ion mode at LHC: 
$\gamma=2950$,
$L=10^{26} cm^{-2} s^{-1}$, Ca-Ca: $\gamma=3750$,
$L=4 \times 10^{30} cm^{-2} s^{-1}$,p-p: $\gamma=7450$,
$L=10^{30} cm^{-2} s^{-1}$.
In the Ca-Ca heavy ion mode, higher 
effective luminosities (defined as collider luminosity times
$\GG$-luminosity) can be achieved as , e.g., in the Pb-Pb mode, 
since higher AA luminosities can be reached there.
For further details see \cite{HenckenTB95}.

\section{$\G$-A interactions}
\label{sec_ga}

There are many interesting phenomena ranging from the excitation of
discrete nuclear states, giant multipole resonances (especially the
giant dipole resonance), quasideuteron absorption, nucleon resonance
excitation to the nucleon continuum.  Photo-induced processes lead in
general to a change of the charge-to-mass ratio of the nuclei, and
with their large cross section they are therefore a serious source of
beam loss. Especially the cross section for the excitation of the
giant dipole resonance, a collective mode of the nucleus, is rather
large for the heavy systems (of the order of 100b).  The cross section
scales approximately with $Z^{10/3}$. Another serious source of beam
loss is the $\EPEM$ bound-free pair creation. The contribution of the
nucleon resonances (especially the $\Delta$ resonance) has also been
confirmed experimentally in fixed target experiments with 60 and~200
GeV/A (heavy ions at CERN, ``electromagnetic spallation'')
\cite{BrechtmannH88a,BrechtmannH88b,PriceGW88}. For details of these
aspects, we refer the reader to
\cite{KraussGS97,VidovicGS93,BaltzRW96,BaurB89}, where scaling laws,
as well as detailed calculations for individual cases are given.

The interaction of quasireal photons with protons has been studied
extensively at the electron-proton collider HERA (DESY, Hamburg), with
$\sqrt{s} = 300$~GeV ($E_e=27.5$~GeV and $E_p=820$~GeV in the
laboratory system). This is made possible by the large flux of
quasi-real photons from the electron (positron) beam. The obtained $\G
p$ center-of-mass energies (up to $W_{\G p}\approx200$~GeV) are an
order of magnitude larger than those reached by fixed target
experiments. Similar and more detailed studies will be possible at the
relativistic heavy ion colliders RHIC and LHC, due to the larger flux
of quasireal photons from one of the colliding nuclei. In the
photon-nucleon subsystem, one can reach invariant masses $W_{\G N}$ up
to $W_{\G N,max}=\sqrt{4 W_{max} E_N} \approx 0.8 \G A^{-1/6}$~GeV. In
the case of RHIC (${}^{197}$Au, $\G=106$) this is about 30~GeV, for
LHC (${}^{208}$Pb, $\G=2950$) one obtains 950~GeV. Thus one can study
physics quite similar to the one at HERA, with nuclei instead of
protons. Photon-nucleon physics includes many aspects, like the energy
dependence of total cross-sections, diffractive and non-diffractive
processes.

An important subject is the elastic vector meson production $\G p
\rightarrow V p$ (with $V=\rho,\omega,\phi,J/\Psi,\dots$). A review of
exclusive neutral vector meson production is given in
\cite{Crittenden97}.  The diffractive production of vector mesons
allows one to get insight into the interface between perturbative QCD
and hadronic physics. Elastic processes (i.e., the proton remains in
the ground state) have to be described within nonperturbative (and
therefore phenomenological) models. It was shown in \cite{RyskinRML97}
that diffractive (``elastic'') $J/\Psi$ photoproduction is a probe of
the gluon density at $x\approx \frac{M_{\Psi}^2}{W_{\G N}^2}$ (for
quasireal photons).  Inelastic $J/\Psi$ photoproduction was also
studied recently at HERA \cite{Breitweg97}.

Going to the hard exclusive photoproduction of heavy mesons on the
other hand, perturbative QCD is applicable. Recent data from HERA on
the photoproduction of $J/\Psi$ mesons have shown a rapid increase of
the total cross section with $W_{\G N}$, as predicted by perturbative
QCD.  Such studies could be extended to photon-nucleus interactions at
RHIC, thus complementing the HERA studies. Equivalent photon flux
factors are large for the heavy ions due to coherence. On the other
hand, the A-A luminosities are quite low, as compared to HERA. Of
special interest is the coupling of the photon of one nucleus to the
Pomeron-field of the other nucleus. Such studies are envisaged for
RHIC, see \cite{KleinS97a,KleinS97b,KleinS95a,KleinS95b} where also
experimental feasibility studies were performed.

Estimates of the order of magnitude of vector meson production in
photon-nucleon processes at RHIC and LHC are given in \cite{BaurHT98}.  In
$AA$ collisions there is incoherent photoproduction on the individual
$A$ nucleons. Shadowing effects will occur in the nuclear environment
and it will be interesting to study these. There is also the coherent
contribution where the nucleus remains in the ground state.  Due to
the large momentum transfer, the total (angle integrated) coherent
scattering shows an undramatic $A^{4/3}$ dependence. (It will be
interesting to study shadow effects in this case also). This is in
contrast to, e.g., low energy $\nu$A elastic scattering, where the
coherence effect leads to an $A^2$ dependence. 
For a general pedagogical discussion of the coherence effects see, e.g., 
\cite{FreedmanST77}. In addition there are inelastic contributions, 
where the proton (nucleon) is transformed into some final state $X$ 
during the interaction (see \cite{Breitweg97}).

At the LHC one can extend these processes to much higher invariant
masses $W$, therefore much smaller values of $x$ will be probed.
Whereas the $J/\Psi$ production at HERA was measured up to invariant
masses of $W\approx 160$~GeV, the energies at the LHC allow for
studies up to $\approx 1$~TeV.

At the LHC \cite{Felix97} hard diffractive vector
meson photoproduction can be investigated especially well in $AA$
collisions. In comparison to previous experiments, the very large
photon luminosity should allow observation of processes with quite
small $\G p$ cross sections, such as $\Upsilon$-production. For more
details see \cite{Felix97}.

\section{Photon-Photon Physics at various invariant mass scales}
\label{chap_proc}

Up to now photon-photon scattering has been mainly studied at $\EPEM$
colliders. Many reviews \cite{BudnevGM75,KolanoskiZ88,BergerW87}
as well as conference reports
\cite{Amiens80,SanDiego92,Sheffield95,Egmond97} exist. The
traditional range of invariant masses has been the region of mesons,
ranging from $\pi^0$ ($m_{\pi^0}=135$~MeV) up to about $\eta_c$
($m_{\eta_c}=2980$~MeV). Recently the total $\GG\rightarrow$~hadron
cross-section has been studied at LEP2 up to an invariant mass range
of about 70~GeV \cite{L3:97}. We are concerned here mainly with the
invariant mass region relevant for RHIC and LHC (see the
$\GG$-luminosity figures below). Apart from the production of $\EPEM$
(and $\MUPM$) pairs, the photons can always be considered as
quasireal. The cross section section for
virtual photons deviates from the one for real photons only for $Q^2$,
which are much larger then the coherence limit $Q^2\lesssim 1/R^2$
(see also the discussion in \cite{BudnevGM75}). For real photons
general symmetry requirements restrict the possible final states, as
is well known from the Landau-Yang theorem. Especially
it is impossible to produce spin 1 final states. In $\EPEM$
annihilation only states with $J^{PC}=1^{--}$ can be produced
directly. Two photon collisions give access to most of the $C=+1$
mesons.

$C=-1$ vector mesons can be produced in principle by the fusion of
three (or, less important, five, seven, \dots) equivalent photons. The
cross section scales with $Z^6$. It is smaller than the contributions
discussed above from $\G A$ collisions, even for nuclei with large $Z$(see
\cite{BaurHT98} 

The cross section for $\GG$-production in a heavy ion collision
factorizes into a $\GG$-luminosity function and a cross-section
$\sigma_{\GG}(W_{\GG})$ for the reaction of the (quasi)real photons
$\GG \rightarrow f$, where $f$ is any final state of interest (see
Eq.~(\ref{eq_sigmaAA}).  When the final state is a narrow resonance,
the cross-section for its production in two-photon collisions is given
by
\BE
\sigma_{\GG\rightarrow R}(M^2) =
 8 \pi^2 (2 J_R+1) \Gamma_{\GG}(R) \delta(M^2-M_R^2)/M_R ,
\label{eq_nres}
\EE
where $J_R$, $M_R$ and $\Gamma_{\GG}(R)$ are the spin, mass and
two-photon width of the resonance $R$. This makes it easy to calculate
the production cross-section $\sigma_{AA\rightarrow AA+R}$ of a
particle in terms of its basic properties.  We will now give a general
discussion of possible photon-photon physics at relativistic heavy ion
colliders. Invariant masses up to several GeV can be reached at RHIC
and up to about 100 GeV at LHC.

We can divide our discussion into the following two main subsections:
Basic QCD phenomena in $\GG$-collisions (covering the range of meson,
meson-pair production, etc.) and $\GG$-collisions as a tool for new
physics, especially at very high invariant masses.
An interesting topic in itself is the $e^+$-$e^-$ pair 
production. The fields are strong enough to produce multiple pairs in a
single collisions. A discussion of this subjet together with calculations 
within the semiclassical approximation can be found in 
\cite{Baur90,HenckenTB95a,HenckenTB95b,alscherHT97}

\subsection{Basic QCD phenomena in $\GG$-collisions}

\subsubsection{Hadron spectroscopy: Light quark spectroscopy}

One may say that photon-photon collisions provide an independent view
of the meson and baryon spectroscopy. They provide powerful
information on both the flavor and spin/angular momentum internal
structure of the mesons. Much has already been done at
$\EPEM$ colliders. Light quark spectroscopy is very
well possible at RHIC, benefiting from the high
$\GG$-luminosities. Detailed feasibility studies exist
\cite{KleinS97a,KleinS97b,KleinS95a,KleinS95b}.  In these studies, $\GG$
signals and backgrounds from grazing nuclear and beam gas collisions
were simulated with both the FRITIOF and VENUS Monte Carlo codes. The
narrow $p_\perp$-spectra of the $\GG$-signals provide a good
discrimination against the background. The possibilities of the LHC
are given in the FELIX LoI \cite{Felix97}.

The absence of meson production via $\GG$-fusion is also of great
interest for glueball search. The two-photon width of a resonance is a
probe of the charge of its constituents, so the magnitude of the
two-photon coupling can serve to distinguish quark dominated
resonances from glue-dominated resonances (``glueballs'').  In
$\GG$-collisions, a glueball can only be produced via the annihilation
of a $q\bar q$ pair into a pair of gluons, whereas a normal $q\bar
q$-meson can be produced directly.
The ``stickiness'' of a mesonic state X is defined as

\BE
S_X = \frac{\Gamma(J/\Psi \rightarrow \G X)}{\Gamma(X \rightarrow
\G \G)} .
\EE
We expect the stickiness of all mesons to be comparable, while for
glueballs it should be enhanced by a factor of about $ 1/\alpha_s^4 \sim 20$.
In a recent reference \cite{Godang97} results of the search for $f_J
(2220)$ production in two-photon interactions were presented. There a
very small upper limit for the product of $\Gamma_{\GG} B_{K_sK_s}$
was given, where $B_{K_s K_s}$ denotes the branching fraction of
its decay into $K_s K_s$.  From this it was concluded that this is a
strong evidence that the $f_J(2220)$ is a glueball.

\subsubsection{Heavy Quark Spectroscopy}

For charmonium production, the two-photon width $\Gamma_{\GG}$ of
$\eta_c$ (2960 MeV, $J^{PC} = 0^{-+}$) is known from experiment. But
the two-photon widths of $P$-wave charmonium states have been measured
with only modest accuracy.  For RHIC the study of $\eta_c$ is a real
challenge \cite{KleinS97b}; the luminosities are falling and the
branching ratios to experimental interesting channels are small.

In Table~\ref{tab_ggmeson} (adapted from table~2.6 of \cite{Felix97})
the two-photon production cross-sections
for $c\bar c$ and $b \bar b$ mesons in the rapidity range $|Y|<7$ are
given. Also given are the number of events in a $10^6$ sec run with
the ion luminosities of $4\times 10^{30}$cm${}^{-2}$s${}^{-1}$ for
Ca-Ca and $10^{26}$cm${}^{-2}$s${}^{-1}$ for Pb-Pb. Millions of
$C$-even charmonium states will be produced in coherent two-photon
processes during a standard $10^6$~sec heavy ion run at the LHC. The
detection efficiency of charmonium events has been estimated as 5\%
for the forward-backward FELIX geometry \cite{Felix97}, i.e., one can
expect detection of about $5\times 10^3$ charmonium events in Pb-Pb
and about $10^6$ events in Ca-Ca collisions. This is two to three
orders of magnitude higher than what is expected during five years of
LEP200 operation. Further details, also on experimental cuts,
backgrounds and the possibilities for the study of $C$-even bottonium
states are given in \cite{Felix97}.
%
%
\begin{table}[hbt]
\begin{center}
\begin{tabular}{|l|c|c|r|c|c|c|}
\hline
State   & Mass,     & $\Gamma_{\GG}$ &
              \multicolumn{2}{|c}{$\sigma (AA\to AA+X)$} &
              \multicolumn{2}{|c|}{Events for $10^6$~sec} \\ 
\cline{4-7}
  & MeV     &  keV  &      Pb-Pb  &                Ca-Ca  &
                           Pb-Pb  &                Ca-Ca  \\
\hline
~~~$\eta'$  & 958 & 4.2 & 22 mb  & 125 $\mu$b  & $2.2 \times 10^7$
                                       & $5.0 \times 10^8$ \\
~~~$\eta_c$ & 2981 & 7.5 & 590 $\mu$b &3.8 $\mu$b &$5.9\times10^5$
                                       & $1.5 \times 10^7$ \\
~~~$\chi_{0c}$& 3415& 3.3& 160 $\mu$b &1.0 $\mu$b &$1.6\times10^5$
                                       & $4.0 \times 10^6$ \\
~~~$\chi_{2c}$& 3556 & 0.8 &160 $\mu$b& 1.0 $\mu$b&$1.6\times10^5$
                                       & $4.0 \times 10^6$ \\
~~~$\eta_b$   & 9366 & 0.43 &370 nb & 3.0 nb   & $370     $
                                              & $12000   $ \\
~~~$\eta_{0b}$& 9860 & $2.5\times10^{-2}$ & 18 nb & 0.14 nb & $18$
                                              & $640     $ \\
~~~$\eta_{2b}$& 9913 & $6.7\times10^{-3}$ & 23 nb & 0.19 nb & $23$
                                              & $76      $ \\
\hline
\end{tabular}
\end{center}
\caption{\it
Production cross sections and event numbers for heavy quarkonia
produced in a $10^6$ sec run in Pb-Pb and Ca-Ca collisions at the LHC
with luminosities
10$^{27}$ and 4$\times 10^{30}~{\rm cm}^{-2} {\rm sec}^{-1}$.
Adapted from \protect\cite{Felix97}.}
\label{tab_ggmeson}
\end{table}

\subsubsection{Vector-meson pair production. Total hadronic
cross-section} 

There are various mechanisms to produce hadrons in photon-photon
collisions. Photons can interact as point particles which produce
quark-antiquark pairs (jets), which subsequently hadronize. Often a
quantum fluctuation transforms the photon into a vector meson
($\rho$,$\omega$,$\phi$, \dots) (VMD component) opening up all the
possibilities of hadronic interactions .  In hard scattering, the
structure of the photon can be resolved into quarks and
gluons. Leaving a spectator jet, the quarks and gluon contained in the
photon will take part in the interaction.  It is of great interest to
study the relative amounts of these components and their properties.

The L3 collaboration recently made a measurement of the total hadron
cross-section for photon-photon collisions in the interval $5 GeV <
W_{\GG} < 75 GeV$ \cite{L3:97}. It was found that the $\GG
\rightarrow$hadrons cross-section is consistent with the universal
Regge behavior of total hadronic cross-sections.
The production of vector meson pairs can well be studied at RHIC with
high statistics in the GeV region \cite{KleinS97a}.  For the
possibilities at LHC, we refer the reader to \cite{Felix97} and 
\cite{BaurHTS98}, where also experimental details and simulations are
described.

\subsection{$\GG$-collisions as a tool for new physics}

The high flux of photons at relativistic heavy ion colliders offers
possibilities for the search of new physics. This includes the
discovery of the Higgs-boson in the $\GG$-production channel or new
physics beyond the standard model, like supersymmetry or
compositeness.

Let us mention here the plans to build an $\EPEM$ linear collider.
Such future linear colliders will be used for $\EPEM$, $e\G$
and $\GG$-collisions (PLC, photon linear collider). 
The photons will be obtained by scattering of
laser photons (of eV energy) on high energy electrons ($\approx$ TeV
region) (see \cite{Telnov95}). Such photons in the TeV energy range
will be monochromatic and polarized. The physics program at such
future machines is discussed in \cite{ginzburg95}, it includes Higgs
boson and gauge boson physics and the discovery of new particles.

While the $\GG$ invariant masses which will be reached at RHIC will
mainly be useful to explore QCD at lower energies, the $\GG$ invariant
mass range at LHC --- up to about 100 GeV --- will open up new
possibilities.

A number of calculations have been made for a medium heavy standard
model Higgs \cite{DreesEZ89,MuellerS90,Papageorgiu95,Norbury90}. For
masses $m_H < 2 m_{W^\pm}$ the Higgs bosons decays dominantly into
$b\bar b$. Chances of finding the standard model
Higgs in this case are marginal \cite{BaurHTS98}.

An alternative scenario with a light Higgs boson was, e.g., given in
\cite{ChoudhuryK97} in the framework of the ``general two Higgs
doublet model''. Such a model allows for a very light particle in the
few GeV region. With a mass of 10~GeV, the $\GG$-width is about 0.1
keV. The authors of
\cite{ChoudhuryK97} proposed to look for such a light neutral Higgs
boson at the proposed low energy $\GG$-collider. We want to point out
that the LHC Ca-Ca heavy ion mode would also be very suitable for such
a search.

One can also speculate about new particles with strong coupling to the
$\GG$-channel. Large $\Gamma_{\GG}$-widths will directly lead to large
$\GG$ production cross-sections. We quote the
papers \cite{Renard83,BaurFF84}. Since the $\GG$-width of a resonance
is mainly proportional to the wave function at the origin, huge values
can be obtained for very tightly bound systems. Composite scalar
bosons at $W_{\GG}\approx 50$~GeV are expected to have $\GG$-widths of
several MeV \cite{Renard83,BaurFF84}. The search for such kind of
resonances in the $\GG$-production channel will be possible at
LHC.

In Refs. \cite{DreesGN94,OhnemusWZ94} $\GG$-processes at $pp$
colliders (LHC) are studied. It is observed there that non-strongly
interacting supersymmetric particles (sleptons, charginos,
neutralinos, and charged Higgs bosons) are difficult to detect in
hadronic collisions at the LHC. The Drell-Yan and gg-fusion mechanisms
yield low production rates for such particles. Therefore the
possibility of producing such particles in $\GG$ interactions at
hadron colliders is examined. Since photons can be emitted from
protons which do not break up in the radiation process, clean events
can be generated which should compensate for the small number. In
\cite{DreesGN94} it was pointed out that at the high luminosity of
$L=10^{34}$cm${}^{-2}$s${}^{-1}$ at the LHC($pp$), one expects about
16 minimum bias events per bunch crossing. Even the elastic $\GG$
events will therefore not be free of hadronic debris. Clean elastic
events will be detectable at luminosities below
$10^{33}$cm${}^{-2}$s${}^{-1}$. This danger of ``overlapping events''
has also to be checked for the heavy ion runs, but it will be much
reduced due to the lower luminosities.

\section{Conclusion}

In this article the basic properties of electromagnetic processes in
very peripheral hadron-hadron collisions are described. The method of
equivalent photons is a well established tool to describe these kinds
of reactions. Reliable results of quasireal photon fluxes and
$\GG$-luminosities are available. Unlike electrons and positrons heavy
ions and protons are particles with an internal structure. Effects
arising from this structure are well under control. A problem, which
is difficult to judge quantitatively at the moment, is the influence
of strong interactions in grazing collisions, i.e., effects arising
from the nuclear stratosphere and Pomeron interactions.

The high photon fluxes open up possibilities for photon-photon as well
as photon-nucleus interaction studies up to energies hitherto
unexplored at the forthcoming colliders RHIC and LHC.  Interesting
physics can be explored at the high invariant $\GG$-masses, where
detecting new particles could be within range. Also very interesting
studies within the standard model, i.e., mainly QCD studies will be
possible. This ranges from the study of the total $\GG$-cross section
into hadronic final states up to invariant masses of about 100~GeV to
the spectroscopy of light and heavy mesons.

\section{Acknowledgement}

One of the authors(G.B.)wishes to thank the organizers for inviting
him to this stimulating conference in a pleasant atmosphere.


\begin{thebibliography}{99}

\bibitem{BertulaniB88}
C.~A. Bertulani and G. Baur, Phys. Rep. {\bf 163},  299  (1988).

\bibitem{BaurR94}
G. Baur and H. Rebel, J. Phys.~G {\bf 20},  1  (1994).

\bibitem{BaurR96}
G. Baur and H. Rebel, Annu. Rev. Nucl. Part. Sci. {\bf 46},  321  (1996).

\bibitem{Primakoff51}
H. Primakoff, Phys. Rev. {\bf 31},  899  (1951).

\bibitem{Moshammer97}
R. Moshammer {\it et~al.}, Phys. Rev. Lett. {\bf 79},  3621  (1997).

\bibitem{LandauL34}
L.~D. Landau and E.~M. Lifshitz, Phys. Z. Sowjet. {\bf 6},  244  (1934).

\bibitem{BaurB88}
G. Baur and C.~A. Bertulani, Z. Phys. A {\bf 330},  77  (1988).

\bibitem{GrabiakMG89}
M. Grabiak, B. M{\"u}ller, W. Greiner, and P. Koch, J. Phys.~G {\bf 15},  L25
  (1989).

\bibitem{Papageorgiu89}
E. Papageorgiu, Phys. Rev.~D {\bf 40},  92  (1989).

\bibitem{Baur90d}
G. Baur,  in {\em CBPF Int. Workshop on relativistic aspects of nuclear
  physics, Rio de Janeiro, Brazil 1989}, edited by T. Kodama {\it et~al.}
  (World Scientific, Singapore, 1990), p.\ 127.

\bibitem{BaurF90}
G. Baur and L.~G. {Ferreira Filho}, Nucl. Phys.~A {\bf 518},  786  (1990).

\bibitem{CahnJ90}
N. Cahn and J.~D. Jackson, Phys. Rev.~D {\bf 42},  3690  (1990).

\bibitem{VidovicGB93}
M. Vidovi{\'c}, M. Greiner, C. Best, and G. Soff, Phys. Rev.~C {\bf 47},  2308
  (1993).

\bibitem{KraussGS97}
F. Krauss, M. Greiner, and G. Soff, Prog. Part. Nucl. Phys. {\bf 39},  503
  (1997).

\bibitem{BaurHT98}
G. Baur, K. Hencken, and D. Trautmann, J. Phys.~G {\bf 24},  1657  (1998).

\bibitem{Vannucci80}
F. Vannucci,  in {\em $\gamma\gamma$ Collisions, Proceedings, Amiens 1980},
  Vol.~134 of {\em Lecture Notes in Physics}, edited by G. Cochard (Springer,
  Heidelberg, Berlin, New York, 1980).

\bibitem{ags98} A. G. Shamov, V. I. Telnov,see the talk by A.Maslennikov 
at this workshop

\bibitem{KleinS97a}
S. Klein and E. Scannapieco,  in {\em Photon '97, Egmond aan Zee}, edited by
  (World Scientific, Singapore, 1997), p.\ 369.

\bibitem{KleinS97b}
S. Klein and E. Scannapieco, Coherent Photons and Pomerons in Heavy Ion
  Collisions, presented at 6th Conference on the Intersections of Particle and
  Nuclear Physics, May 1997, Big Sky, Montana, STAR Note 298, LBNL-40495, 1997.

\bibitem{KleinS95a}
S. Klein and E. Scannapieco, STAR Note 243, 1995.

\bibitem{KleinS95b}
S. Klein,  in {\em Photon '95, Sheffield}, edited by D.~J. Miller, S.~L.
  Cartwright, and V. Khoze (World Scientific, Singapore, 1995), p.\ 417.

\bibitem{jny98} J. Nystrand, invited talk, this conference.

\bibitem{HenckenKKS96}
{K. Hencken, {Yu.} V. Kharlov, G. V. Khaustov, S. A. Sadovsky, and V. D.
  Samoylenko}, TPHIC, event generator of two photon interactions in heavy ion
  collisions, IHEP-96-38, 1996.

\bibitem{Felix97}
K. Eggert {\it et~al.}, FELIX Letter of Intent, CERN/LHCC 97--45, LHCC/I10,
  1997.

\bibitem{BaurHTS98}
{G. Baur, K. Hencken, D. Trautmann, S. Sadovsky, and Yu. Kharlov},
  Photon-Photon Physics with heavy ions at CMS, CMS Note 1998/009, available
  from the CMS information server at http://cmsserver.cern.ch, 1998.

\bibitem{Bjorken99}
J.~D. Bjorken, Nucl. Phys. B (Proc. Suppl.) {\bf 71},  484  (1999).

\bibitem{BudnevGM75}
V.~M. Budnev, I.~F. Ginzburg, G.~V. Meledin, and V.~G. Serbo, Phys. Rep. {\bf
  15},  181  (1975).

\bibitem{JacksonED}
J.~D. Jackson, {\em Classical Electrodynamics} (John Wiley, New York, 1975).

\bibitem{WintherA79}
A. Winther and K. Alder, Nucl. Phys.~A {\bf 319},  518  (1979).

\bibitem{BaurF91}
G. Baur and L.~G. {Ferreira Filho}, Phys. Lett.~B {\bf 254},  30  (1991).

\bibitem{BaurB93}
G. Baur and N. Baron, Nucl. Phys.~A {\bf 561},  628  (1993).

\bibitem{Baur92}
G. Baur, Z. Phys. C {\bf 54},  419  (1992).

\bibitem{HenckenTB95}
K. Hencken, D. Trautmann, and G. Baur, Z. Phys. C {\bf 68},  473  (1995).

\bibitem{BrechtmannH88a}
C. Brechtmann and W. Heinrich, Z. Phys. A {\bf 330},  407  (1988).

\bibitem{BrechtmannH88b}
C. Brechtmann and W. Heinrich, Z. Phys. A {\bf 331},  463  (1988).

\bibitem{PriceGW88}
P.~B. Price, R. Guaxiao, and W.~T. Williams, Phys. Rev. Lett. {\bf 61},  2193
  (1988).

\bibitem{VidovicGS93}
M. Vidovi{\'c}, M. Greiner, and G. Soff, Phys. Rev.~C {\bf 48},  2011  (1993).

\bibitem{BaltzRW96}
A.~J. Baltz, M.~J. Rhoades-Brown, and J. Weneser, Phys. Rev. E {\bf 54},  4233
  (1996).

\bibitem{BaurB89}
G. Baur and C.~A. Bertulani, Nucl. Phys.~A {\bf 505},  835  (1989).

\bibitem{Crittenden97}
J.~A. Crittenden, {\em Exclusive production of neutral vector mesons at the
  electron-proton collider HERA}, Vol.~140 of {\em Springer tracts in modern
  physics} (Springer, Heidelberg, 1997).

\bibitem{RyskinRML97}
M.~G. Ryskin, R.~G. Roberts, A.~D. Martin, and E.~M. Levin, Z. Phys. C {\bf
  76},  231  (1997).

\bibitem{Breitweg97}
J. Breitweg {\it et~al.}, Z. Phys. C {\bf 76},  599  (1997).

\bibitem{FreedmanST77}
D.~Z. Freedman, D.~N. Schramm, and D.~L. Tubbs, Annu. Rev. Nucl. Part. Sci.
  {\bf 27},  167  (1977).

\bibitem{KolanoskiZ88}
H. Kolanoski and P. Zerwas,  in {\em High Energy Electron-Positron Physics},
  edited by A. Ali and P. S{\"o}ding (World Scientific, Singapore, 1988).

\bibitem{BergerW87}
{Ch. Berger and W. Wagner}, Phys. Rep. {\bf 176C},  1  (2987).

\bibitem{Amiens80}
{\em {$\gamma\gamma$} Collisions, Proceedings, Amiens 1980}, Vol.~134 of {\em
  Lecture Notes in Physics}, edited by G. Cochard and P. Kessler (Springer,
  Berlin, 1980).

\bibitem{SanDiego92}
{\em Proc. 9th International Workshop on Photon-Photon Collisions, San Diego
  (1992)} (World Scientific, Singapore, 1992).

\bibitem{Sheffield95}
{\em Photon'95, Xth International Workshop on Gamma-Gamma Collisions and
  related Processes}, edited by D.~J. Miller, S.~L. Cartwright, and V. Khoze
  (World Scientific, Singapore, 1995).

\bibitem{Egmond97}
{\em Photon'97, XIth International Workshop on Gamma-Gamma Collisions and
  related Processes, Egmond aan Zee}, edited by A. Buijs (World Scientific,
  Singapore, 1997).

\bibitem{L3:97}
{L3 collaboration}, Phys. Lett.~B {\bf 408},  450  (1997).

\bibitem{Baur90}
G. Baur, Phys. Rev.~A {\bf 42},  5736  (1990).

\bibitem{HenckenTB95a}
K. Hencken, D. Trautmann, and G. Baur, Phys. Rev.~A {\bf 51},  998  (1995).

\bibitem{HenckenTB95b}
K. Hencken, D. Trautmann, and G. Baur, Phys. Rev.~A {\bf 51},  1874  (1995).

\bibitem{alscherHT97}
A. Alscher, K. Hencken, D. Trautmann, and G. Baur, Phys. Rev.~A {\bf 55},  396
  (1997).

\bibitem{Godang97}
R. Godang {\it et~al.}, Phys. Rev. Lett. {\bf 79},  3829  (1997).

\bibitem{Telnov95}
V. Telnov,  in {\em Photon '95, Sheffield}, edited by D.~J. Miller, S.~L.
  Cartwright, and V. Khoze (World Scientific, Singapore, 1995), p.\ 369.

\bibitem{ginzburg95}
I.~F. Ginzburg,  in {\em Photon '95, Sheffield}, edited by D.~J. Miller, S.~L.
  Cartwright, and V. Khoze (World Scientific, Singapore, 1995), p.\ 399.

\bibitem{DreesEZ89}
M. Drees, H. Ellis, and D. Zeppenfeld, Phys. Lett.~B {\bf 223},  454  (1989).

\bibitem{MuellerS90}
B. {M\"uller} and A.~J. Schramm, Phys. Rev.~D {\bf 42},  3699  (1990).

\bibitem{Papageorgiu95}
E. Papageorgiu, Phys. Lett.~B {\bf 352},  394  (1995).

\bibitem{Norbury90}
J. Norbury, Phys. Rev.~D {\bf 42},  3696  (1990).

\bibitem{ChoudhuryK97}
D. Choudhury and M. Krawczyk, Phys. Rev.~D {\bf 55},  2774  (1997).

\bibitem{Renard83}
F.~M. Renard, Phys. Lett.~B {\bf 126},  59  (1983).

\bibitem{BaurFF84}
U. Baur, H. Fritzsch, and H. Faissner, Phys. Lett.~B {\bf 135},  313  (1984).

\bibitem{DreesGN94}
M. Drees, R.~M. Godbole, N. Nowakowski, and S.~D. Rindami, Phys. Rev.~D {\bf
  50},  2335  (1994).

\bibitem{OhnemusWZ94}
J. Ohnemus, T.~F. Walsh, and P.~M. Zerwas, Phys. Lett.~B {\bf 328},  369
  (1994).

\end{thebibliography}

\end{document}